\documentclass[3p]{elsarticle}
\usepackage{amsmath}

\usepackage{amssymb}

\usepackage[usenames,dvipsnames]{xcolor}
\usepackage[inline]{enumitem}
\usepackage{hyperref}
\newcommand\currentversion{2019.0}
\newcommand\bcalV{\boldsymbol{\cal V}}
\newcommand{\MOLSCAT}{{\sc molscat}}
\newcommand{\FIELD}{{\sc field}}

\newcommand{\etal}{{\textit {et al.{}}}}
\newcommand\inpitem[1]{{\tt #1}}
\newcommand\basisitem[1]{{\tt #1}}

\newcommand\namelist[1]{{\tt #1}}

\newcommand\file[1]{{\tt #1}}
\newcommand\prog[1]{{\tt #1}}
\newcommand\mylabel[1]{\label{#1}
}
\colorlet{crls}{blue}
\colorlet{jmh}{red}
\newcounter{bla}

\journal{Computer Physics Communications}

\begin{document}

\begin{frontmatter}



\title{\MOLSCAT: a program for non-reactive quantum scattering calculations \\
on atomic and molecular collisions}


\author{Jeremy M. Hutson\corref{jmh}}
\author{C. Ruth Le Sueur\corref{}}

\cortext[jmh]{Corresponding author. \textit{E-mail address:} J.M.Hutson@durham.ac.uk}
\address{Joint Quantum Centre (JQC)
Durham-Newcastle, Department of Chemistry, \\ University of Durham, South Road,
Durham, DH1 3LE, UK}

\begin{abstract}
\MOLSCAT\ is a general-purpose program for quantum-mechanical calculations on
nonreactive atom-atom, atom-molecule and molecule-molecule collisions. It
constructs the coupled-channel equations of atomic and molecular scattering
theory, and solves them by propagating the wavefunction or log-derivative
matrix outwards from short range to the asymptotic region at long range. It
then applies scattering boundary conditions to extract the scattering matrix (S
matrix). Built-in coupling cases include atom + rigid linear molecule, atom +
vibrating diatom, atom + rigid symmetric top, atom + asymmetric or spherical
top, rigid diatom + rigid diatom, rigid diatom + asymmetric top, and
diffractive scattering of an atom from a crystal surface. Interaction
potentials may be specified either in program input (for simple cases) or with
user-supplied routines. For the built-in coupling cases, \MOLSCAT\ can loop
over total angular momentum (partial wave) and total parity to calculate
elastic and inelastic integral cross sections and spectroscopic line-shape
cross sections. Post-processors are available to calculate differential cross
sections, transport, relaxation and Senftleben-Beenakker cross sections, and to
fit the parameters of scattering resonances.  \MOLSCAT\ also provides an
interface for plug-in routines to specify coupling cases (Hamiltonians and
basis sets) that are not built in; plug-in routines are supplied to handle
collisions of a pair of alkali-metal atoms with hyperfine structure in an
applied magnetic field. For low-energy scattering, \MOLSCAT\ can calculate
scattering lengths and effective ranges and can locate and characterise
scattering resonances as a function of an external variable such as the
magnetic field.
\end{abstract}


\end{frontmatter}


\noindent {\bf PROGRAM SUMMARY}

\begin{small}
\noindent {\em Manuscript Title: }\MOLSCAT: a program for non-reactive quantum
scattering calculations on atomic and molecular collisions \\
{\em Authors: }Jeremy M. Hutson and C. Ruth Le Sueur  \\
{\em Program Title: }\MOLSCAT \\
{\em Journal Reference:} \\
{\em Catalogue identifier:} \\
{\em Licensing provisions: }none  \\
{\em Programming language: }Fortran 90 \\
{\em Computer:}                                               \\
{\em Operating system:}                                       \\
{\em RAM:} case dependent                                     
{\em Classification:} 16.7 Elastic Scattering and Energy Transfer \\
{\em External routines/libraries:} LAPACK, BLAS               \\
{\em Subprograms used:}                                       \\
\\
{\em Nature of problem:}
\\
Quantum-mechanical calculations of scattering properties for non-reactive
collisions between atoms and molecules.
\\
{\em Solution method:}
\\
The Schr\"odinger equation is expressed in terms of coupled equations in the
interparticle distance, $R$. Solutions of the coupled-channel equations are
propagated outwards from the classically forbidden region at short range to the
asymptotic region. The program calculates scattering S matrices and uses them
to calculate scattering properties including elastic, inelastic and line-shape
cross sections.
\\
\\
{\em Unusual features: }
\begin{enumerate}
\item \MOLSCAT\ contains numerous features to handle quantities important in
    low-energy collisions. It can propagate very efficiently to very long
    range, and it can calculate low-energy properties such as the
    scattering length (complex and energy-dependent if required) and the
    effective range.
\item \MOLSCAT\ can construct and solve sets of coupled equations using basis
    sets that are not eigenfunctions of the Hamiltonians of the colliding
    particles at long range. The propagation is carried out in the
    primitive basis set, and the solutions are transformed to the
    asymptotic basis set before applying long-range boundary conditions to
    extract the S matrix.
\item \MOLSCAT\ provides an interface that allows users to specify the coupled
    equations that arise from additional Hamiltonians and basis sets.
\item \MOLSCAT\ provides an interface that allows users to include multiple
    external fields in the Hamiltonian.  This same interface also allows
    users to include a factor which scales the interaction potential, and
    to investigate how properties vary with this factor.
\item \MOLSCAT\ can locate and characterise low-energy Feshbach resonances
    either as a function of collision energy or as a function of external
    fields or of the potential scaling factor.
\end{enumerate}
{\em Running time:}
Highly dependent on mass and complexity of interacting particles\\
\end{small}

\section{Introduction}
\label{Theory}

\MOLSCAT\ solves the quantum-mechanical non-reactive scattering problem for
systems where the total Hamiltonian of the colliding pair may be written
\begin{equation}
H=-\frac{\hbar^2}{2\mu}R^{-1}\frac{d^2\,}{d R^2}R
+\frac{\hbar^2 \hat L^2}{2\mu R^2}+H_{\rm intl}(\xi_{\rm intl})+V(R,\xi_{\rm intl}),
\label{eqh}
\end{equation}
where $R$ is a radial coordinate describing the separation of two particles and
$\xi_{\rm intl}$ represents all the other coordinates in the system. $H_{\rm
intl}$ represents the sum of the internal Hamiltonians of the isolated
particles, and depends on $\xi_{\rm intl}$ but not $R$, and $V(R,\xi_{\rm
intl})$ is an interaction potential. The operator $\hbar^2 \hat L^2/2\mu R^2$
is the centrifugal term that describes the end-over-end rotational energy of
the interacting pair.

Scattering problems of this type commonly arise in atomic and molecular
collisions~\cite{Bernstein:1979}. The observable quantities are often
differential or integral cross sections for elastic and inelastic collisions
processes, or the corresponding thermally averaged rate coefficients. There are
also cross sections that describe relaxation properties and spectroscopic line
shapes (line widths, line shifts and line mixing)~\cite{Shafer:1973} and
transport properties and Senftleben-Beenakker effects~\cite{Liu:1979}.
Scattering calculations are also used to model spectroscopic line positions,
widths and lifetimes for predissociating states of van der Waals
complexes~\cite{Ashton:1983}. Somewhat different applications arise for
ultracold atomic
and molecular collisions~\cite{Hutson:theory-cold-colls:2009}, where the
quantities of interest are often scattering lengths, total inelastic loss
rates, and the properties of zero-energy Feshbach resonances.

The internal Hamiltonian $H_{\rm intl}$ is a sum of terms for the two particles
1 and 2,
\begin{equation}
H_{\rm intl}(\xi_{\rm intl}) = H_{\rm intl}^{(1)}(\xi_{\rm intl}^{(1)})
+ H_{\rm intl}^{(2)}(\xi_{\rm intl}^{(2)}),
\end{equation}
with eigenvalues $E_{{\rm intl},i}=E_{{\rm intl},i}^{(1)}+E_{{\rm
intl},i}^{(2)}$, where $E_{{\rm intl},i}^{(1)}$ and $E_{{\rm intl},i}^{(2)}$
are energies of the separated monomers $1$ and $2$. The individual terms can
vary enormously in complexity: each one may represent a structureless atom,
requiring no internal Hamiltonian at all, a vibrating and/or rotating molecule,
or a particle with electron and/or nuclear spins. The problems that arise in
ultracold physics frequently involve pairs of atoms or molecules with electron
and nuclear spins, often in the presence of external electric, magnetic or
photon fields. All these complications can be taken into account in the
structure of $H_{\rm intl}$ and the interaction potential $V(R,\xi_{\rm
intl})$, which may both involve terms dependent on spins and external fields.

\MOLSCAT\ solves the Schr\"o\-ding\-er equation for the Hamiltonian \eqref{eqh}
using the \emph{coupled-channel} approach, which handles the radial coordinate
$R$ by direct numerical propagation on a grid, and all the other coordinates
using a basis set. In the coupled-channel approach, the total wavefunction is
expanded
\begin{equation} \Psi(R,\xi_{\rm intl})
=R^{-1}\sum_j\Phi_j(\xi_{\rm intl})\psi_{j}(R), \label{eqexp}
\end{equation}
where the functions $\Phi_j(\xi_{\rm intl})$ form a complete orthonormal basis
set for motion in the coordinates $\xi_{\rm intl}$ and the factor $R^{-1}$
serves to simplify the form of the radial kinetic energy operator. The
wavefunction in each {\em channel} $j$ is described by a radial \emph{channel
function} $\psi_{j}(R)$. The expansion (\ref{eqexp}) is substituted into the
total Schr\"odinger equation, and the result is projected onto a basis function
$\Phi_i(\xi_{\rm intl})$. The resulting coupled differential equations for the
channel functions $\psi_{i}(R)$ are
\begin{equation}\frac{d^2\psi_{i}}{d R^2}
=\sum_j\left[W_{ij}(R)-{\cal E}\delta_{ij}\right]\psi_{j}(R).
\end{equation}
Here $\delta_{ij}$ is the Kronecker delta and ${\cal E}=2\mu E/\hbar^2$, where
$E$ is the total energy, and
\begin{equation}
W_{ij}(R)=\frac{2\mu}{\hbar^2}\int\Phi_i^*(\xi_{\rm intl}) [\hbar^2 \hat L^2/2\mu R^2 +
H_{\rm intl}+V(R,\xi_{\rm intl})] \Phi_j(\xi_{\rm intl})\,d\xi_{\rm intl}. \label{eqWij}
\end{equation}
The different equations are coupled by the off-diagonal terms $W_{ij}(R)$ with
$i\ne j$.

The coupled equations may be expressed in matrix notation,
\begin{equation}
\frac{d^2\boldsymbol{\psi}}{d R^2}= \left[{\bf W}(R)-{\cal E}{\bf
I}\right]\boldsymbol{\psi}(R). \label{eqcp}
\end{equation}
If there are $N$ basis functions included in the expansion (\ref{eqexp}),
$\boldsymbol{\psi}(R)$ is a column vector of order $N$ with elements
$\psi_{j}(R)$, ${\bf I}$ is the $N\times N$ unit matrix, and ${\bf W}(R)$ is an
$N\times N$ interaction matrix with elements $W_{ij}(R)$.

In general there are $N$ linearly independent solution vectors
$\boldsymbol{\psi}(R)$ that satisfy the Schr\"o\-ding\-er equation subject to
the boundary condition that $\boldsymbol{\psi}(R)\rightarrow0$ in the
classically forbidden region at short range. These $N$ column vectors form a
wavefunction matrix $\boldsymbol{\Psi}(R)$. The various propagators in
\MOLSCAT\ work either by propagating $\boldsymbol{\Psi}(R)$ and its radial
derivative $\boldsymbol{\Psi}'(R)$ or by propagating the log-derivative matrix
${\bf Y}(R)=\boldsymbol{\Psi}'(R)[\boldsymbol{\Psi}(R)]^{-1}$.

The particular choice of the basis functions $\Phi_j(\xi_{\rm intl})$ and the
resulting form of the interaction matrix elements $W_{ij}(R)$ depend on the
physical problem being considered. The complete set of coupled equations often
factorises into blocks determined by the symmetry of the system. In the absence
of external fields, the \emph{total angular momentum} $J_{\rm tot}$ and the
\emph{total parity} are conserved quantities. Different or additional
symmetries arise in different physical situations. \MOLSCAT\ is designed to
loop over total angular momentum and parity, constructing a separate set of
coupled equations for each combination and solving them by propagation. These
loops may be repurposed for other symmetries when appropriate.

\MOLSCAT\ can also handle interactions that occur in external fields, where the
total angular momentum is no longer a good quantum number.

\subsection{Matrix of the interaction potential}\label{theory:W}

In order to streamline the calculation of matrix elements for the propagation,
\MOLSCAT\ expresses the interaction potential in an expansion over the internal
coordinates,
\begin{equation}
V(R,\xi_{\rm intl})=\sum_\Lambda v_\Lambda(R){\cal V}^\Lambda(\xi_{\rm intl}).
\label{eqvlambda}
\end{equation}
The specific form of the expansion depends on the nature of the colliding
particles. The radial potential coefficients $v_\Lambda(R)$ may either be
supplied explicitly, or generated internally by numerically integrating over
$\xi_{\rm intl}$. The $R$-independent coupling matrices $\bcalV^\Lambda$ with
elements ${\cal V}^\Lambda_{ij}=\langle\Phi_i|{\cal
V}^\Lambda|\Phi_j\rangle_{\rm intl}$ are calculated once and stored for use in
evaluating $W_{ij}(R)$ throughout the course of a propagation.

\subsection{Matrices of the internal and centrifugal Hamiltonians}\label{theory:Wextra}

Coupled-channel scattering theory is most commonly formulated in a basis set
where $\hat L^2$ and $H_{\rm intl}$ are both diagonal. All the built-in
coupling cases use basis sets of this type. The matrix of $H_{\rm intl}$ is
$\langle\Phi_i|H_{\rm intl}|\Phi_j\rangle_{\rm intl}=E_{{\rm
intl},i}\delta_{ij}$. The diagonal matrix elements of $\hat L^2$ are often of
the form $L_i(L_i+1)$, where the integer quantum number $L_i$ (sometimes called
the partial-wave quantum number) represents the end-over-end angular momentum
of the two particles about one another.

However, \MOLSCAT\ also allow the use of basis sets where one or both of $\hat
L^2$ and $H_{\rm intl}$ are non-diagonal. If $H_{\rm intl}$ is non-diagonal, it
is expanded as a sum of terms
\begin{equation}
H_{\rm intl}(\xi_{\rm intl})
=\sum_\Omega h_\Omega {\cal H}^\Omega_{\rm intl}(\xi_{\rm intl}),
\label{eqHomega1}
\end{equation}
where the $h_\Omega$ are scalar quantities, some of which may represent
external fields if desired. \MOLSCAT\ generates additional coupling matrices
$\boldsymbol{\cal H}^\Omega$ with elements ${\cal
H}^\Omega_{ij}=\langle\Phi_i|{\cal H}^\Omega_{\rm intl}|\Phi_j\rangle_{\rm
intl}.$ These are also calculated once and stored for use in evaluating
$W_{ij}(R)$ throughout the course of a propagation. A similar mechanism is used
for basis sets where $\hat L^2$ is non-diagonal, with
\begin{equation}
\hat L^2
=\sum_\Upsilon {\cal L}^\Upsilon.
\label{eqL2}
\end{equation}

If $H_{\rm intl}$ is non-diagonal, the allowed energies $E_{{\rm intl},i}$ of
the pair of monomers at infinite separation are the eigenvalues of $H_{\rm
intl}$. The wavefunctions of the separated pair are represented by simultaneous
eigenvectors of $H_{\rm intl}$ and $\hat L^2$.

\subsection{Boundary conditions at short range}

For scattering calculations where the origin is energetically inaccessible, it
is often sufficient to require that $\boldsymbol{\Psi}(R)\rightarrow0$ in the
classically forbidden region at short range, or equivalently that ${\bf
Y}(R)\rightarrow\pm\infty$. However, it is usually more efficient to apply a
boundary condition based on the Wentzel-Kramers-Brillouin (WKB) approximation,
\begin{eqnarray}
Y(R) \approx k(R),\label{eq:bcwkb}
\end{eqnarray}
where $k(R) = [2\mu(V(R)-E)/\hbar^2]^{1/2}$ and $V(R)$ is an effective
potential energy for the channel concerned. For locally closed channels,
\MOLSCAT\ applies the boundary conditions (\ref{eq:bcwkb}) by default. This
gives faster convergence with respect to $R_{\rm min}$ than ${\bf
Y}(R)\rightarrow \infty$.

There are also circumstances where more general boundary conditions are
required:
\begin{itemize}[nosep]
\item In systems where $R=0$ is energetically accessible, some states
    require ${\bf Y}(0)=0$.
\item In model systems with a Fermi pseudopotential, corresponding to a
    $\delta$-function at the origin or elsewhere, a finite value of ${\bf
    Y}$ may be required.
\end{itemize}
\MOLSCAT\ allows the imposition of separate boundary conditions for ${\bf Y}$
in locally closed and open channels at $R_{\rm min}$.

\subsection{The scattering matrix}\label{theory:scatcalcs}

The outcome of a collision process is usually described in quantum mechanics by
the scattering matrix (S matrix), which contains information on the probability
amplitudes and phases for the various possible outcomes. In simple cases
(diagonal $H_{\rm intl}$ and $\hat L^2$), each possible outcome corresponds to
one of the channels in the coupled equations. Alternatively, if $H_{\rm intl}$
is non-diagonal, each outcome corresponds to an \emph{asymptotic channel}
represented by one of the simultaneous eigenvectors of $H_{\rm intl}$ and $\hat
L^2$ with energy eigenvalue $E_{{\rm intl},i}$. Each asymptotic channel $i$ is
{\em open} if it is energetically accessible as $R\rightarrow\infty$ ($E_{{\rm
intl},i}\le E$) or \emph{closed} if it is energetically forbidden ($E_{{\rm
intl},i}>E$).

For each $J_{\rm tot}$ and symmetry block, solutions to the coupled equations
are propagated from deep inside the classically forbidden region at short range
to a distance at long range beyond which the interaction potential may be
neglected. The wavefunction matrix $\boldsymbol{\Psi}(R)$ and its radial
derivative (or the log-derivative matrix ${\bf Y}(R)$) are then matched to the
analytic functions that describe the solutions of the Schr\"odinger equation in
the absence of an interaction potential,
\begin{equation}
\boldsymbol{\Psi}(R)={\bf J}(R)+{\bf N}(R){\bf K}.
\end{equation}
where the matrices ${\bf J}(R)$ and ${\bf N}(R)$ are diagonal and are made up
of Ricatti-Bessel functions for the open channels and modified spherical Bessel
functions for the closed channels. For each channel $i$, the Bessel function is
of order $L_i$ and its argument is $k_i R$, where $k_i$ is the asymptotic
wavevector such that $\hbar^2 k_i^2/2\mu=|E-E_{{\rm intl},i}|$. For some basis
sets, $L_i$ is non-integer.

The real symmetric $N\times N$ matrix ${\bf K}$ is then converted to the S
matrix,
\begin{equation}
{\bf S}=({\bf I}+i{\bf K}_{\rm oo})^{-1}({\bf I}-i{\bf K}_{\rm oo}),
\end{equation}
where ${\bf K}_{\rm oo}$ is the open-open portion of ${\bf K}$. ${\bf S}$ is
a complex symmetric unitary matrix of dimension $N_{\rm open}\times N_{\rm
open}$, where $N_{\rm open}$ is the number of open channels.

If $\hat L^2$ and $H_{\rm intl}$ are both diagonal, the asymptotic channels
used for matching to Bessel functions are the same as the channels used to
propagate the wavefunction matrix $\boldsymbol{\Psi}(R)$ or its log-derivative
${\bf Y}(R)$.

If $\hat L^2$ and/or $H_{\rm intl}$ is non-diagonal, \MOLSCAT\ transforms
$\boldsymbol{\Psi}(R)$ or ${\bf Y}(R)$ at $R=R_{\rm max}$ into a basis set that
diagonalises $\hat L^2$ and $H_{\rm intl}$:
\begin{itemize}[leftmargin=13pt]
\item If $\hat L^2$ is diagonal but $H_{\rm intl}$ is not, \MOLSCAT\
    constructs the matrix of $H_{\rm intl}$ for each value of $L$ in turn,
    and diagonalises it. It uses the resulting eigenvectors to transform
    the corresponding block of $\boldsymbol{\Psi}(R_{\rm max})$ or ${\bf
    Y}(R_{\rm max})$ into the asymptotic basis set.
\item If $\hat L^2$ is non-diagonal (whether $H_{\rm intl}$ is diagonal or
    not), \MOLSCAT\ constructs the complete matrix of $H_{\rm intl}$ and
    diagonalises it. If there are degenerate eigenvalues $E_{{\rm intl},i}$
    of $H_{\rm intl}$, it then constructs the matrix of $\hat L^2$ for each
    degenerate subspace and diagonalises it to obtain eigenvalues $L_i$ and
    simultaneous eigenvectors of $\hat L^2$ and $H_{\rm intl}$. It uses the
    simultaneous eigenvectors to transform $\boldsymbol{\Psi}(R_{\rm max})$
    or ${\bf Y}(R_{\rm max})$ into the asymptotic basis set.
\end{itemize}
Finally \MOLSCAT\ uses the transformed $\boldsymbol{\Psi}(R_{\rm max})$ or
${\bf Y}(R_{\rm max})$, together with the eigenvalues $E_{{\rm intl},i}$ and
$L_i$, to extract ${\bf K}$ and ${\bf S}$.

There are occasional cases where there are sets of vectors that are degenerate
or near-degenerate in both $\hat L^2$ and $H_{\rm intl}$. It is possible to
define extra operators $\hat P_i$ to aid in resolving such
\mbox{(near-)degeneracies}. These extra operators, which need not necessarily
contribute to $H_{\rm intl}$ or $\hat L^2$, must nevertheless commute with
them. In any \mbox{(near-)degenerate} subspace of the eigenvectors of $H_{\rm
intl}$ and $\hat L^2$, \MOLSCAT\ constructs the matrix of the first such
operator ($\hat P_1$) and finds linear combinations of the degenerate functions
that are eigenfunctions of it. If the eigenvalues of $\hat P_1$ are
sufficiently non-degenerate, the process ends; if not, it is repeated with
operator $\hat P_2$, and so on.

\subsection{Cross sections}\label{theory:cross-sec}

Experimental observables that describe completed collisions, such as
differential and integral cross sections, and scattering lengths, can be
written in terms of S-matrix elements. Cross sections typically involve a
\emph{partial-wave sum}, with contributions from many values of $J_{\rm tot}$,
except at the lowest kinetic energies (in the ultracold regime). By default
\MOLSCAT\ uses its S matrices to accumulate degeneracy-averaged state-to-state
integral cross sections, which may be written
\begin{equation}
\sigma_{n_{\rm i}\rightarrow n_{\rm f}} = \frac{\pi}{g_{n_{\rm i}} k_{n_{\rm i}}^2}
\sum_{\substack{J_{\rm tot}\\M}} (2J_{\rm tot}+1)
\sum_{\substack{i\in n_{\rm i}\\f\in n_{\rm f}}}
\left|\delta_{if}-S_{if}^{J_{\rm tot},M}\right|^2.\label{eqsigdef}
\end{equation}
Here $n_{\rm i}$ and $n_{\rm f}$ label initial and final levels (not states) of
the colliding pair,\footnote{For most coupling cases, $n_{\rm i}$ and $n_{\rm
f}$ each represent several quantum numbers, not just one.} while $i$ and $f$
indicate the open channels arising from those levels for total angular momentum
$J_{\rm tot}$ and symmetry block $M$. $g_{n_{\rm i}}$ is the degeneracy of
level $n_{\rm i}$.

\MOLSCAT\ can calculate line-shape cross sections for the broadening, shifting
and mixing of spectroscopic lines for most of the built-in coupling cases.
Line-shape cross sections require scattering calculations for the upper and
lower states of the spectroscopic transition at the same \emph{kinetic} (not
total) energy; if desired, the input energies are interpreted as kinetic
energies and the total energies required are generated internally.

\MOLSCAT\ has features to locate scattering resonances, which may produce sharp
features in the energy-dependence of cross sections and may also be interpreted
as predissociating states of van der Waals complexes. It can calculate the
S-matrix eigenphase sum, which is a generalisation of the scattering phase
shift to multichannel problems~\cite{Ashton:1983}, and use it to converge
towards a far-off scattering resonance as a function of energy.

\MOLSCAT\ can output S matrices to auxiliary files for later processing.
Separate programs are available:
\begin{itemize}[nosep]
\item program \prog{DCS}~\cite{DCS} to calculate differential cross
    sections;
\item program \prog{SBE}~\cite{SBE} to calculate generalised transport,
    relaxation and Senft\-leben-Beenakker cross sections;
\item program \prog{RESFIT}~\cite{Hutson:resfit:2007} to fit to eigenphase
    sums and S-matrix elements to extract resonance positions, widths and
    partial widths as a function of energy or external field.
\end{itemize}

\subsection{Reference energy}

By default, the zero of energy used for total energies is the one used for
monomer energies, or defined by the monomer Hamiltonians programmed in a
plug-in basis-set suite. However, it is sometimes desirable to use a different
zero of energy (reference energy). This may be specified:
\begin{itemize}[nosep]
\item{as a value given directly in the input file;}
\item{as the energy of a particular scattering threshold or pair of monomer
    states, which may depend on external fields.}
\end{itemize}

\subsection{Low-energy collision properties}\mylabel{theory:lowE}

\MOLSCAT\ has many features designed to facilitate low-energy scattering
calculations.

\subsubsection{Scattering lengths and volumes}

\MOLSCAT\ can calculate scattering lengths or volumes, which may be complex in
the presence of inelastic channels. The diagonal S-matrix element in an
incoming channel 0 may be written in terms of a complex phase shift $\eta$,
\begin{equation}
S_{00}=\exp(2i\eta).
\end{equation}
For a channel with low kinetic energy, the phase shift may be expanded in
powers of the incoming wavevector $k$. The leading term is proportional to $k$
for $s$-wave scattering and $k^3$ for $p$-wave scattering. For higher values of
$L$ the leading power depends on the form of the long-range potential: e.g.,
for a potential that varies as $-C_6R^{-6}$ at long range, the leading term is
proportional to $k^4$ for all $L>1$. \MOLSCAT\ calculates scattering lengths
for $L=0$ ($n=1$), volumes for $L=1$ ($n=3$) and hypervolumes for $L>1$ ($n=4$)
using the formula~\cite{Hutson:res:2007}
\begin{equation}
a_L(k)=\frac{-\tan\eta}{k^n}=\frac{1}{ik^n}\left(\frac{1-S_{00}}{1+S_{00}}\right).
\label{eq:scatln}\end{equation}
It should be emphasised that Eq.~\ref{eq:scatln} is an identity, so that this
is a far more general approach than that used by some other programs that take
the limit of other (often more complicated) functions as $k\rightarrow0$. The
scattering length and volume defined by Eq.~\ref{eq:scatln} become independent
of $k$ at sufficiently low $k$.

\subsubsection{Characterisation of zero-energy Feshbach resonances}

\MOLSCAT\ can converge on and characterise the zero-energy Feshbach resonances
that appear in the scattering length as a function of external fields~\cite{Frye:resonance:2017}.
It can do this for resonances both in elastic
scattering, where the scattering length has a simple pole characterised by its
position and width and the background scattering length, and in inelastic
scattering, where the resonant behaviour is more complex and requires
additional parameters~\cite{Hutson:res:2007}.

It is often convenient to locate resonances using the associated \FIELD\
program \cite{bound+field:2019} before characterising them with \MOLSCAT.

\subsubsection{Effective range}

In the absence of inelastic scattering, the $s$-wave scattering length $a_0$ is
real. The near-threshold dependence of the $s$-wave scattering phase shift
$\eta$ on kinetic energy $E_{\rm kin}$ or wavevector $k$ is often characterised
by an effective-range expansion at small collision momentum
\begin{equation}
k\cot\eta(k)=-\frac{1}{a_0(0)}+\frac{1}{2} r_{\rm eff}k^2+\cdots
\end{equation}
Either the scattering length $a_0(k)$ or its reciprocal can thus be expressed
at low energy in terms of the effective range $r_{\rm eff}$~\cite{Blackley:eff-range:2014}.
The first of these is numerically unstable near
a pole and the second near a zero crossing. \MOLSCAT\ calculates the effective
range using both expansions for every energy after the first.

\subsection{Scattering wavefunctions}

If desired, \MOLSCAT\ can calculate the multichannel scattering wavefunction
$\boldsymbol{\psi}(R)$ that is incoming in a single channel and outgoing in all
channels, using a generalisation of the method of Hutson and Thornley~\cite{THORNLEY:1994}.

\subsection{Infinite-order sudden approximation}\label{theory:IOS}

\MOLSCAT\ incorporates code for calculating degeneracy-averaged and line-shape
cross sections within the infinite-order sudden (IOS) ansatz. In this
formulation, S matrices are calculated from propagations carried out at fixed
molecular orientations. The cross sections are written in terms of sums of
products of dynamical factors $Q$ and spectroscopic coefficients $F$. The
dynamical factors contain all the information about the collision dynamics;
they are defined as integrals over the fixed-orientation S matrices, which are
evaluated by numerical quadrature. The spectroscopic coefficients contain
information about rotor levels and angular momentum coupling. The IOS code has
not been used much in recent years, but is retained for backwards
compatibility. It does not provide low-energy features such as scattering
lengths.

\section{Systems handled}\label{interactiontypes}

\MOLSCAT\ can carry out scattering calculations in the close-coupling
approximation (with no dynamical approximations except basis-set truncation)
for the following pairs of species:
\begin{enumerate}[nosep]
\item Atom + linear rigid rotor~\cite{Arthurs:1960};
\item Atom + vibrating diatom (rotationally and/or vibrationally inelastic)
    with interaction potentials independent of diatom rotational
    state~\cite{Green:1979:vibrational};
\item Linear rigid rotor + linear rigid
    rotor~\cite{Green:1975,Green:1977:comment,Heil:1978:coupled};
\item Asymmetric top + linear molecule~\cite{Phillips:1995}
\item Atom + symmetric top (also handles near-symmetric tops and linear
    molecules with vibrational angular
    momentum)~\cite{Green:1976,Green:1979:IOS};
\item Atom + asymmetric top~\cite{Green:1976} (also handles spherical
    tops~\cite{Hutson:spher:1994});
\item Atom + vibrating diatom (rotationally and/or vibrationally inelastic)
    with interaction potentials dependent on diatom rotational
    state~\cite{Hutson:sbe:1984};
\item Atom + rigid corrugated surface: diffractive (elastic)
    scattering~\cite{Wolken:1973:surface,Hutson:1983}. At present, the code
    is restricted to centrosymmetric lattices, for which the potential
    matrices are real.
\end{enumerate}
The close-coupling calculations are all implemented in a fully coupled
space-fixed representation, with the calculations performed separately for each
total angular momentum and parity.

In addition, \MOLSCAT\ implements a variety of dynamical approximations
(decoupling methods) that offer considerable savings of computer time at the
expense of accuracy. The methods supported are:
\begin{enumerate}[nosep]
\item{Effective potential approximation~\cite{Rabitz:EP} (seldom used
    nowadays);}
\item{Coupled-states (centrifugal sudden) approximation~\cite{McG74};}
\item{Decoupled $L$-dominant
    approximation~\cite{Green:1976:DLD,DePristo:1976:DLD} (seldom used
    nowadays).}
\end{enumerate}
Not all these approximations are supported for all collision types, though the
most common ones are.

In addition to the built-in coupling cases, \MOLSCAT\ provides an interface
that allows users to specify sets of coupled equations that arise for different
pairs of colliding species. These have been used for numerous different cases,
and routines are provided for two cases of current interest:
\begin{enumerate}[nosep]
\item Structureless atom + $^3\Sigma$ molecule in a magnetic field,
    demonstrated for Mg + NH;
\item Two alkali-metal atoms in $^2$S states, including hyperfine coupling,
    in a magnetic field, demonstrated for $^{85}$Rb$_2$.
\end{enumerate}

\section{Propagators}\label{propagators}

\MOLSCAT\ can solve the coupled equations using a variety of different
propagation methods. The propagators currently implemented are:

\begin{itemize}[leftmargin=13pt]
\item{Log-derivative propagator of Johnson (LDJ)~\cite{Johnson:1973,
    Manolopoulos:1993:Johnson}: This is a very stable propagator. It has
    largely been superseded by the LDMD propagator.}
\item{Diabatic log-derivative propagator of Manolopoulos
    (LDMD)~\cite{Manolopoulos:1986}: This is a very efficient and stable
    propagator, especially at short and medium range. It is coded to detect
    single-channel cases (including IOS cases) automatically and in that
    case use a more efficient implementation.}
\item{Quasiadiabatic log-derivative propagator of Manolopoulos
    (LDMA)~\cite{Manolopoulos:PhD:1988, Hutson:CPC:1994}: This is similar
    to the LDMD propagator, but operates in a quasiadiabatic basis. It
    offers better accuracy than LDMD for very strongly coupled problems,
    but is relatively expensive. It is recommended for production runs only
    for very strongly coupled problems. However, it is also useful when
    setting up a new system, because it can output eigenvalues of the
    interaction matrix at specific distances (adiabats) and nonadiabatic
    couplings between the adiabatic states.}
\item{Symplectic log-derivative propagators of Manolopoulos and Gray (LDMG)
    \cite{MG:symplectic:1995}: This offers a choice of 4th-order or
    5th-order symplectic propagators. These are 1.5 to 3 times more
    expensive per step than the LDMD and LDJ propagators, but can have
    smaller errors for a given step size.  They can be the most efficient
    choice when high precision is required.}
\item{Airy propagator: This is the AIRY log-derivative propagator of
    Alexander~\cite{Alexander:1984} as reformulated by Alexander and
    Manolopoulos~\cite{Alexander:1987}.  It uses a quasiadiabatic basis
    with a linear reference potential (which results in Airy functions as
    reference solutions). This allows the step size to increase rapidly
    with separation, so that this propagator is particularly efficient at
    long range.}
\item{de Vogelaere propagator (DV) \cite{deVogelaere:1955}: This propagates
    the wavefunction explicitly, but is much slower than more modern
    methods, especially for large reduced masses or high scattering
    energies. It is not recommended except for special purposes.}
\item{R-matrix propagator (RMAT) \cite{Stechel:1978}: This is a stable
    method that works in a quasiadiabatic basis. It has relatively poor
    step-size convergence properties, and has largely been superseded by
    the log-derivative propagators. It is not recommended except for
    special purposes.}
\item{VIVS propagator: This is the variable-interval variable-step method
    of Parker \etal\ \cite{Parker:VIVS} and is intended for use at long
    range. It is sometimes very efficient, but the interval size is limited
    when there are deeply closed channels, so that it is not efficient at
    long range in such cases. Control of it is considerably more
    complicated than for other propagators, and it has largely been
    superseded by the AIRY propagator.}
\item{WKB semiclassical integration using Gauss-Mehler
    quadrature~\cite{Pac74}: This is not a true propagator and can be used
    only for single-channel problems.}
\end{itemize}

\MOLSCAT\ can use different propagators at short and long range. This is
particularly useful for low-energy scattering, where it may be necessary to
propagate to very large values of $R$ to obtain converged results. The AIRY
propagator incorporates a variable step size, and can be used to propagate
outwards with a fast-increasing step size at long range at very low cost.
However, it is not particularly efficient when the interaction potential is
fast-varying, so it is often used in combination with a fixed-step-size method
such as the LDMD propagator at short and intermediate range.

\section{Computer time}

The computer time required to solve a set of $N$ coupled equations is
approximately proportional to $N^3$. The practical limit on $N$ is from a few
hundred to several thousand, depending on the speed of the computer and the
amount of memory available.

The computer time also depends linearly on the number of radial steps required
to solve the coupled equations to the desired accuracy. The step size required
is typically proportional to the minimum local wavelength, so that the time
scales approximately with $(\mu E_{\rm max}^{\rm kin})^{1/2}$, where $E_{\rm
max}^{\rm kin}$ is the maximum local kinetic energy; for scattering
calculations, $E_{\rm max}^{\rm kin}$ may be approximated by the sum of the
collision energy and the well depth of the interaction potential.

\section{Plug-in functionality}

\subsection{Potential or potential expansion coefficients}

The programs internally express the interaction potential as an expansion over
the internal coordinates, as in Eq.~\eqref{eqvlambda}. The expansion
coefficients $v_\Lambda(R)$ may be supplied in a variety of ways:
\begin{itemize}[nosep, leftmargin=13pt]
\item For very simple potentials, where the functions $v_\Lambda(R)$ are
    sums of exponentials and inverse powers, the parameters that specify
    them may be supplied in the input file.
\item For more complicated functions, plug-in routines may be supplied to
    return individual values of $v_\Lambda(R)$ at a value of $R$ specified
    in the calling sequence.
\item For most of the built-in coupling cases, plug-in routines may be
    supplied to return the unexpanded potential $V(R,\xi_{\rm intl})$ at
    specified values of $R$ and the internal coordinates $\xi_{\rm intl}$.
    The general-purpose potential routine supplied then performs numerical
    quadrature over $\xi_{\rm intl}$ to evaluate the expansion coefficients
    $v_\Lambda(R)$ internally.
\item If none of these approaches is convenient (or efficient enough), a
    replacement potential routine may be supplied to return the complete
    potential expansion at once (all values of $v_\Lambda(R)$ at a value of
    $R$ specified in the calling sequence).
\end{itemize}

\subsection{Basis sets and coupling matrices}

The programs provide an interface for users to supply a set of routines that
specify an additional type of basis set, select the elements that will be used
in a particular calculation, and calculate the matrices of coupling
coefficients for the operators ${\cal V}^\Lambda(\xi_{\rm intl})$ used to
expand the interaction potential. The routines must also specify the matrices
of $H_{\rm intl}$ and $\hat L^2$, which may be diagonal or non-diagonal. If
desired, $H_{\rm intl}$ may contain terms that depend on external fields.

\subsection{External fields and potential scaling}

The programs incorporate data structures to handle external electric, magnetic
or photon fields. There may be multiple fields at arbitrary relative
orientations. Internally, the field-dependent terms in the Hamiltonian are a
subset of those in $H_{\rm intl}$,
\begin{equation}
H_{\rm intl}(\xi_{\rm intl},\boldsymbol{B})
=\sum_\Omega B_\Omega {\cal H}^\Omega_{\rm intl}(\xi_{\rm intl}),
\label{eqHomegaB}
\end{equation}
where the vector $\boldsymbol{B}$ represents all the fields present. The
elements of $\boldsymbol{B}$ may each be expressed as a \emph{nonlinear}
function of external field variables (EFVs); the EFVs may thus (for example)
represent the magnitudes, orientations, or relative angles of the component
fields. \MOLSCAT\ allows calculations on a grid of values of any one EFV, and
allows the location and characterisation of low-energy Feshbach resonances as a
function of one EFV with all the others fixed.

The programs also allow calculations as a function of a scaling factor that
multiplies the entire interaction potential, or a subset of the potential
expansion coefficients $v_\Lambda(R)$. The scaling factor is handled internally
using the same structures as external fields.

\section{Distributed files and example calculations}\label{use:input}

\subsection{Distributed files}

The program is supplied as a tarred zipped file, which contains:
\begin{itemize}
\item{the full program documentation in pdf format;}
\item{a directory {\tt source\_code} containing
\begin{itemize}
\item{the Fortran source code;}
\item{a GNU makefile ({\tt GNUmakefile}) that can build the executables
    needed for the example calculations;}
\end{itemize}}
\item{a directory {\tt examples} containing
\begin{itemize}
\item{a sub-directory {\tt input} containing input files for the
    example calculations described below;}
\item{a sub-directory {\tt output} containing the corresponding output files;}
\end{itemize}}
\item{a directory {\tt data} containing auxiliary data files
    for some potential routines used in the example calculations;}
\item{a plain-text file {\tt README} that gives information on changes that
    may be needed to adapt the GNUmakefile to a specific target computer.}
\item{a plain-text file {\tt COPYING} that contains the text of the GNU General Public License, Version 3.}
\end{itemize}

\subsection{Example calculations}

The executables used for different calculations may differ in the routines
linked to construct the basis set, specify the internal Hamiltonian, and
evaluate the interaction potential. The executables required for the example
calculations can all be built using {\tt GNUmakefile}.

\subsubsection{All available propagators}
\label{testfiles:molscat:intflgs}

\begin{tabular}{ll}
input file: & \file{molscat-all\_propagators.input}\\
executable: & \file{molscat-basic}
\end{tabular}

\file{molscat-all\_propagators.input} performs close-coupling calculations for
a single partial wave and parity for a simple model of collisions between an
atom and a linear rigid rotor and prints the resulting S matrix. The radial
potential coefficients are provided in the input file and consist of a
Lennard-Jones 12-6 potential for $\lambda=0$ and a dispersion-like $R^{-6}$
form for $\lambda=2$. The calculation is repeated using combinations of
short-range and long-range propagators that exercise every propagation method
available in \MOLSCAT\ (though not every possible combination).

\subsubsection{All available decoupling approximations}\label{testfiles:molscat:iadds}

\begin{tabular}{ll}
input file: & \file{molscat-all\_iadds.input}\\
executable: & \file{molscat-basic}
\end{tabular}

\file{molscat-all\_iadds.input} contains input data for a similar model system,
extended this time to include vibrations of the linear rotor. The radial
potential coefficients are again provided in the input file, and all consist of
inverse-power functions of $R$.  The LDMD/AIRY hybrid propagation scheme is
used. \MOLSCAT\ first performs close-coupling calculations and then repeats the
calculation using every decoupling approximation available.

\subsubsection{Location of quasibound state (Feshbach resonance) for Ar + HF}\label{testfiles:molscat:ityp1}

\begin{tabular}{ll}
input file: & \file{molscat-Ar\_HF.input}\\
executable: & \file{molscat-Rg\_HX}\\
also required:& \file{data/h2even.dat}
\end{tabular}

\file{molscat-Ar\_HF.input} demonstrates the procedure for locating a narrow
resonance in the S-matrix eigenphase sum as a function of energy. It performs
calculations on the H6(4,3,2) potential of Hutson~\cite{H92ArHF} for the ground
($v=0$) vibrational state of HF, using the LDMD propagator. It first does
scattering calculations at 5 energies reasonably close to the resonance (but
actually over 1000 widths away) and uses the resulting eigenphase sums to
estimate the resonance position. It then does another 5 calculations around the
estimated resonance position, and this time finds that they span the resonance
(which is between points 2 and 3 of the second set of 5). The formula used for
estimating resonance positions is valid only far from resonance, so it reports
that the second set of points cannot safely be used to locate the resonance
energy.

The extrapolation procedure used in this example amplifies any tiny differences
between computers due to finite-precision arithmetic. The second set of 5
energies is commonly significantly different on different computers; this does
not indicate an error.

\subsubsection{Line-shape cross sections for Ar + CO$_2$}\label{testfiles:molscat:ityp1b}

\begin{tabular}{ll}
input file: & \file{molscat-Ar\_CO2.input}\\
executable: & \file{molscat-Rg\_CO2}
\end{tabular}

\file{molscat-Ar\_CO2.input} performs close-coupling calculations of line-shape
cross sections for the S(10) Raman line of CO$_2$ in Ar, using the single
repulsion potential of Hutson \etal~\cite{H96ArCO2fit} with the LDMD propagator
at short range and the AIRY propagator at long range. The calculations are at a
kinetic energy of 200 cm$^{-1}$ and the total energies are calculated
internally. The program prints cross sections accumulated up to the current
value of $J_{\rm TOT}$; the convergence of the partial-wave sum may be compared
with Fig.\ 2 of ref.\ \cite{Roc97CO2scat}.

\subsubsection{Line-shape cross sections for Ar + H$_2$}\label{testfiles:molscat:ityp7}

\begin{tabular}{ll}
input file: & \file{molscat-Ar\_H2.input}\\
executable: & \file{molscat-Rg\_H2}\\
also required: & \file{data/h2even.dat}
\end{tabular}

\file{molscat-Ar\_H2.input} calculates pure rotational Raman line widths and
shifts across a shape resonance at a collision energy near 14 cm$^{-1}$. It
uses the BC$_3$(6,8) interaction potential of Carley and Le Roy~\cite{RJL80},
evaluated for H$_2$ states $(j,v) = (0,0)$, (2,0) and (4,0) using H$_2$ matrix
elements in the file \file{data/h2even.dat}. The line-shape calculations
require S matrices evaluated at the same \emph{kinetic} energy for different
rotational states of H$_2$; the program treats the input energies as kinetic
energies and generates the total energies required. The results may be compared
with Figure~2(a) of ref.~\cite{Hutson:sbe:1984}.

\subsubsection{Cross sections for rigid rotor + rigid rotor collisions}
\label{testfiles:molscat:ityp3}

\begin{tabular}{ll}
input file: & \file{molscat-ityp3.input}\\
executable: & \file{molscat-H2\_H2}
\end{tabular}

\file{molscat-ityp3.input} contains input data for collisions between
para-H$_2$ and ortho-H$_2$. The interaction potential is that of Zarur and
Rabitz~\cite{Zarur:1974}.

\MOLSCAT\ calculates elastic and state-to-state inelastic cross sections. This
calculation requires that contributions from partial waves are accumulated
until the partial-wave sums are converged within the limits set by the input
data. Two pairs of calculations are performed, first with close-coupling
calculations and then with the coupled-states approximation. All calculations
use the LDMD/AIRY hybrid propagation scheme.

Each calculation is done twice; one in which the radial potential coefficients
are provided, and one in which the unexpanded potential is provided and is
expanded by quadrature by the program. The two calculations illustrate the
equivalence of the two methods.

\subsubsection{Cross sections for atom + symmetric top collisions, with
automated testing of propagator convergence} \label{testfiles:molscat:ityp5}

\begin{tabular}{ll}
input file: & \file{molscat-ityp5.input}\\
executable: & \file{molscat-basic}
\end{tabular}

\file{molscat-ityp5.input} contains input data for atom + symmetric top
collisions between He and ortho-NH$_3$, taking account of the tunnelling
splitting of NH$_3$. It uses a simple analytical interaction potential and the
hybrid propagation scheme. The input file uses $\basisitem{ISYM}(3)=1$ to
select rotational functions of E symmetry and $\basisitem{ISYM}(4)=1$ to
specify that the H nuclei are fermions. The first four calculations are for a
single partial wave and exercise the convergence-testing code in \MOLSCAT,
testing the convergence with respect to step size (chosen in two ways), and
with respect to the start point and end point of the propagation. The final two
calculations carry out full cross-section calculations, using converged values
for the propagation variables, first with close-coupling calculations and then
with the coupled-states approximation.

\subsubsection{Cross sections for atom + spherical top collisions}
\label{testfiles:molscat:ityp6}

\begin{tabular}{ll}
input file: & \file{molscat-ityp6.input}\\
executable: & \file{molscat-Ar\_CH4}
\end{tabular}

\file{molscat-ityp6.input} contains input data for atom + spherical top
collisions between Ar and CH$_4$, using the interaction potential of Buck
\etal~\cite{Buck:1983}. The ground-state rotational constants and the
tetrahedral centrifugal distortion constant $d_{\rm t}$ are specified in the
input file and the program uses them to calculate properly symmetrised
spherical-top wavefunctions. The input file selects CH$_4$ rotor functions of A
symmetry by setting \basisitem{ISYM} to 224. The cross sections use the
automatic total angular momentum option $\inpitem{JTOTU}=99999$ with a
convergence tolerance (\inpitem{OTOL}) of 0.0001 to give well-converged
inelastic cross sections, but the diagonal convergence tolerance \inpitem{DTOL}
is set to 10.0 so that the partial-wave sum terminates before the elastic cross
sections are converged. The results may be compared with Table~VI of
ref.~\cite{Chapman:1996}, although they do not agree exactly because the
results in the paper are averaged over the experimental distribution of
collision energies.

\subsubsection{Atom-surface scattering}\label{testfiles:molscat:ityp8}

\begin{tabular}{ll}
input file: & \file{molscat-ityp8.input}\\
executable: & \file{molscat-basic}
\end{tabular}

\file{molscat-ityp8.input} contains input data for diffractive scattering
($\basisitem{ITYPE}=8$) of He from solid LiF, using the model potential of
Wolken~\cite{Wolken:1973:surface}. It uses the LDMD propagator at two energies.

\subsubsection{Cross sections for Mg + NH in a magnetic field}\label{testfiles:molscat:MgNH}

\begin{tabular}{ll}
input file: & \file{molscat-Mg\_NH.input}\\
executable: & \file{molscat-Mg\_NH}\\
also required: & \file{data/pot-Mg\_NH.data}
\end{tabular}

\file{molscat-Mg\_NH.input} contains input data for cold collisions of NH with
Mg in a magnetic field. It uses a plug-in basis-set suite for a $^3\Sigma$
diatom colliding with a structureless atom. Radial potential coefficients are
obtained by RKHS interpolation using the interaction potential of Sold\'an
\etal~\cite{Soldan:MgNH:2009}.  The coupled equations are solved using the
LDMD/AIRY hybrid propagation scheme.

The basis-set suite implements two different forms of the monomer Hamiltonian,
including and excluding the off-diagonal matrix elements of the spin-spin
operator. The input file specifies runs with both of these. The approximation
does not actually produce any saving in computer time in this case.

The input file requests calculations at kinetic energies of 1, 10 and 100~mK
above the $n=0$, $j=1$, $m_j=1$ threshold of NH.

These calculations are similar to (a subset of) those of Wallis \etal~\cite{Wallis:MgNH:2009},
although the test run uses a smaller basis set than
ref.~\cite{Wallis:MgNH:2009}. Convergence at 100~mK requires inclusion of
incoming partial waves up to $L=3$, which requires values of $M_{\rm tot}$ from
$-2$ to 4 for incoming $m_j=1$.

\MOLSCAT\ can accumulate cross sections from calculations for different values
of $M_{\rm tot}$ and parity at a single field. The first part of the test run,
including the off-diagonal matrix elements of the spin-spin operator,
illustrates the scheme used for identification of levels for systems with
non-diagonal Hamiltonians, where not all threshold channels may be known at the
point where the first partial cross sections are calculated.

\subsubsection{Characterisation of magnetically tunable Feshbach resonances
and calculation of effective range for $^{85}$Rb + $^{85}$Rb}
\label{testfiles:molscat:85Rb2}

\begin{tabular}{ll}
input file: & \file{molscat-Rb2.input}\\
executable: & \file{molscat-Rb2}
\end{tabular}

\file{molscat-Rb2.input} contains input data for low-energy $^{85}$Rb+$^{85}$Rb
collisions in a magnetic field, using a plug-in basis-set suite for a pair of
alkali-metal atoms in a magnetic field, including hyperfine interactions. It
uses the potential of Strauss \etal~\cite{Strauss:2010}, implemented with
potential coefficients incorporated in the executable.  All the calculations
use the LDMD/AIRY hybrid propagation scheme.

This test run characterises 4 different low-energy Feshbach resonances as a
function of magnetic field, using the characterisation algorithms described by
Frye and Hutson~\cite{Frye:resonance:2017}. The first resonance is in purely
elastic scattering in the lowest (aa) scattering channel, so produces a pole in
the scattering length as a function of magnetic field. The second and third
resonances occur in collisions at excited thresholds, where weak inelastic
scattering is possible and the pole is replaced by an oscillation~\cite{Hutson:res:2007}.
The fourth is subject to strong background
inelasticity. Following those 4 calculations, a further calculation obtains the
effective range across the strong resonance observed in the aa channel near
850~G, from scans across the resonance at energies of 100 and 200 nK.

The basis-set suite for this interaction requires information about the
hyperfine properties of the atoms in an additional namelist block named
\namelist{\&BASIS9}. The potential
expansion comprises 3 terms: the singlet and triplet interaction potentials,
and the spin-spin dipolar term, which is modelled in the form
\begin{equation}
\lambda(R)=E_{\rm h}\alpha^2\left[\frac{g_S^2}{4(R/a_0)^3}+A\exp(-\beta R/a_0)\right].
\end{equation}

\section{Program history}\label{history}

The \MOLSCAT\ program was originally written by Sheldon Green in the 1970s,
incorporating propagators from several different authors and adapting them to
share the same input / output structures and mechanisms for generating matrix
elements of the molecular Hamiltonian and interaction potential. Early versions
of the program handled atom-molecule and molecule-molecule scattering, with a
variety of additional coupling cases added between versions~1 (1973) and 7
(1979)~\cite{molscat:NRCC:1980}. Both full space-fixed close-coupling
calculations and decoupling approximations such as coupled states  /
centrifugal-sudden (CS), infinite-order sudden (IOS) and decoupled $L$-dominant
(DLD) approximations were implemented. The program calculated integral elastic,
state-to-state inelastic and line-shape cross sections internally and wrote S
matrices to a file that could be used for post-processors, including
\prog{DCS}~\cite{DCS} for differential cross sections and \prog{SBE}~\cite{SBE} for cross
sections associated with transport and relaxation properties and
Senftleben-Beenakker effects.

Jeremy Hutson became a co-author of \MOLSCAT\ in 1982 and collaborated with
Sheldon Green on its development until Green's death in 1995.  Major changes in
this period included new coupling cases, including the diffractive scattering
of atoms from crystal surfaces, the replacement of some older propagators with
new ones, and code to handle scattering resonances (with post-processor
\prog{RESFIT}~\cite{Hutson:resfit:2007} for fitting resonance parameters). In
addition, an interface was added in version 11 (1992) to allow the inclusion of
plug-in coupling cases with angular momentum algebra and/or interaction
potential expansions that were not already present in the code.

The last version of \MOLSCAT\ produced by Green and Hutson in collaboration was
version 14 in 1994. This was distributed via the CCP6 collaboration in the
UK~\cite{molscat:v14} and via the NASA GISS website. Version 14 formed the
basis of a parallel version named {{\scshape PMP}} \MOLSCAT, produced by George
McBane~\cite{McBane:PMP:2005}.

There has been no fully documented publication of the program since \MOLSCAT\
version 14.

\subsection{Principal changes in version \currentversion}\label{changes}

\begin{itemize}[leftmargin=13pt]
\item The basis-set plug-in mechanism has been extended to allow
    propagation in basis sets that are not eigenfunctions of the internal
    Hamiltonian $H_{\rm intl}$ and/or $\hat L^2$. This makes implementing
    new types of system much simpler than before, especially where the
    individual interaction partners have complicated Hamiltonians.

\item If $H_{\rm intl}$ or $\hat L^2$ is non-diagonal, the propagated
    wavefunction or log-derivative matrix is transformed into a basis set
    that is diagonal in both operators before matching to long-range
    functions to extract the S matrix.

\item The basis-set plug-in functionality has been used to add new
    capabilities to carry out calculations in external fields (electric,
    magnetic, and/or photon) and to loop over (sets of) values of the
    fields.

\item The S matrix is now processed to calculate scattering lengths (or
    volumes or hypervolumes) $a_L$ [actually $k$-dependent complex
    scattering lengths or volumes $a_L(k)$] for any low-energy scattering
    channels. \MOLSCAT\ can also extrapolate the real part of $a_0(k)$ to
    $k=0$ and calculate the effective range $r_{\rm eff}$.

\item \MOLSCAT\ can now converge on and characterise Feshbach resonances as
    a function of external field. This is implemented for both the elastic
    case, where resonances appear as poles in the scattering length/volume,
    and the inelastic case, where the resulting decayed resonances have
    more complicated signatures.

\item \MOLSCAT\ can now calculate a multichannel scattering wavefunction
    that is incoming in a single channel and outgoing in all open channels.

\item A more general mechanism for combining propagators for use at short and long
range has been implemented, which allows any sensible combination.

\item A more general choice of log-derivative boundary conditions at the starting
point for propagation is now allowed.

\item An additional propagation approach~\cite{MG:symplectic:1995} has been
    included, implemented by George McBane, which takes advantage of the
    symplectic nature of the multi-channel radial Schr\"odinger equation.
\end{itemize}

\section{Acknowledgements}

We are grateful to an enormous number of people who have contributed routines,
ideas, and comments over the years. Any attempt to list them is bound to be
incomplete. Many of the early contributors are mentioned in the program history
in section \ref{history}. In particular, we owe an enormous debt to the late
Sheldon Green, who developed the original \MOLSCAT\ program and established
many structures that have proved general enough to support the numerous later
developments. He also developed the \prog{DCS} and \prog{SBE} post-processors.
Robert Johnson, David Manolopoulos, Millard Alexander, Gregory Parker and
George McBane all contributed propagation methods and routines. Christopher
Ashton added code to calculate eigenphase sums and developed the \prog{RESFIT}
post-processor. Timothy Phillips developed code for interactions between
asymmetric tops and linear molecules. Alice Thornley developed methods to
calculate bound-state wavefunctions from log-derivative propagators and George
McBane extended them to scattering wavefunctions. Maykel Leonardo
Gonz\'alez-Mart\'\i{}nez worked on the addition of structures for non-diagonal
Hamiltonians, including magnetic fields, and Matthew Frye contributed
algorithms for converging on low-energy Feshbach resonances (both elastic and
inelastic).

This work was supported by the U.K. Engineering and Physical Sciences Research
Council (EPSRC) Grant Nos.\ EP/P01058X/1, EP/P008275/1 and EP/N007085/1.





\bibliographystyle{elsarticle-num}
\bibliography{../all}







\end{document}